\documentclass[aps,prd,showpacs,
twocolumn
]{revtex4}
\usepackage{graphicx}
\usepackage{hyperref}
\usepackage{amssymb, amsmath}

\def\la{\; \raise0.3ex\hbox{$<$\kern-0.75em\raise-1.1ex\hbox{$\sim$}}\;}
\def\ga{\;  \raise0.3ex\hbox{$>$\kern-0.75em\raise-1.1ex\hbox{$\sim$}}\;}

\def\pFn{p_{\raise-0.3ex\hbox{{\scriptsize F$\!$\raise-0.03ex\hbox{\rm n}}}}
}  
\def\pFp{p_{\raise-0.3ex\hbox{{\scriptsize F$\!$\raise-0.03ex\hbox{\rm p}}}}
}  
\def\pFe{p_{\raise-0.3ex\hbox{{\scriptsize F$\!$\raise-0.03ex\hbox{\rm e}}}}
}  
\def\pFmu{p_{\raise-0.3ex\hbox{{\scriptsize F$\!$\raise-0.03ex\hbox{
\rm $\mu$}}}} }  
\def\m@th{\mathsurround=0pt }
\def\eqalign#1{\null\,\vcenter{\openup1\jot \m@th
   \ialign{\strut$\displaystyle{##}$&$\displaystyle{{}##}$\hfil
   \crcr#1\crcr}}\,}
\def\dd{\mbox{d}}

\begin{document}

\title{Bulk viscosity of superfluid neutron stars}
%
\author{Mikhail E. Gusakov}
\affiliation{
Ioffe Physical Technical Institute,
Politekhnicheskaya 26, 194021 Saint-Petersburg, Russia
}
\date{}
%


\begin{abstract}
The hydrodynamics, describing dynamical effects 
in superfluid neutron stars,
essentially differs from the standard 
one-fluid hydrodynamics.
In particular, we have four bulk viscosity coefficients 
in the theory instead of one.
In this paper we calculate 
these coefficients, for the first time, 
assuming they are due to non-equilibrium beta-processes
(such as modified or direct Urca process).
The results of our analysis are used to estimate 
characteristic damping times 
of sound waves in superfluid neutron stars.
It is demonstrated that all four bulk viscosity coefficients 
lead to comparable dissipation of sound waves
and should be considered on the same footing.
\end{abstract}

\maketitle

\section{Introduction}
\label{1}

The matter in pulsating neutron stars
is not (even locally) in chemical equilibrium.
Particles of different kinds turn into one another
so that the system evolves to equilibrium.
If a deviation from the equilibrium is small then
the processes of mutual transformations of particles can
be described in terms of an effective bulk viscosity 
(see, e.g., Ref.\ \cite{ll87}).
This viscosity influences the `instability windows',
that are the regions of physical parameters 
(e.g., the rotation period and temperature of a star)
at which the neutron star becomes unstable against the 
radiation of gravitational waves
\cite{Lindblom95,ak01,andersson03,andersson06}.
The bulk viscosity, generated by non-equilibrium processes 
of particle transformations, 
was calculated in a series of papers
for neutron-star matter of various composition 
(for example, for matter composed of neutrons, protons, and electrons 
with an admixture of muons; for hyperon or quark matter).
A short review and references to these papers 
can be found in Ref.\ \cite{dsw07}.

It is generally agreed that the stellar matter becomes superfluid 
at a certain stage of neutron star thermal evolution
\cite{yls99,ls01,yp04}.
A lot of attention has been paid
to the question 
how superfluidity affects the bulk viscosity
(see, e.g., Refs.\ \cite{hly00,hly01,hly02,lo02,no06,ssr06}).
In analogy with the ordinary hydrodynamics of non-superfluid liquid,
the only one `standard' bulk viscosity coefficient 
has been calculated and analyzed in all these papers. 
Meanwhile, it is well known 
\cite{putterman74,ll87,khalatnikov89}
that a
superfluid liquid, 
composed of identical particles, 
is generally described by 
the {\it four} bulk viscosity coefficients.
So, what can be expected from neutron stars, 
which contain a mixture of many superfluid species?

In this paper we show that non-equilibrium processes
of particle transformations lead to the appearance
of at least four bulk viscosity coefficients.
Each of them is important for analyzing 
dissipative processes in superfluid neutron stars.  
To be specific, we consider the simplest model
of stellar matter composed of 
neutrons, protons, and electrons (npe-matter).
In this case the bulk viscosity is 
associated with
the non-equilibrium direct or modified Urca process.

The paper is organized as follows. 
In Sec.\ \ref{2} we phenomenologically obtain 
the general form of the dissipative corrections to the 
equations of relativistic hydrodynamics \cite{Son01,ga06}, 
describing a superfluid liquid composed of identical particles. 
In Sec.\ \ref{3} this dissipative hydrodynamics is 
generalized to describe superfluid mixtures
and applied to npe-matter.
In Sec.\ \ref{4} we calculate and analyse 
all four bulk viscosity coefficients provided by
non-equilibrium beta-processes.
For illustration of these results,
in Sec.\ \ref{5} we calculate the characteristic damping times 
of sound waves in superfluid npe-matter.
Sec.\ \ref{6} presents summary.

\section{The dissipative relativistic hydrodynamics 
of one-component superfluid liquid}
\label{2}
In this section we obtain the general form of dissipative terms 
entering the equations 
of relativistic superfluid hydrodynamics
of electrically neutral liquid composed of identical particles.
For that purpose, we need to choose 
a version of non-dissipative hydrodynamics.
There is a number of equivalent formulations
of non-dissipative relativistic superfluid hydrodynamics 
\cite{Carter76,Carter79,Carter85,kl82,lk82,Son01}.
The most elegant and general amongst them seems to be the 
formulation of Carter \cite{Carter76,Carter79,Carter85} 
in which the hydrodynamic
equations follow from a convective variational principle.
Most of the relativistic calculations 
(see Refs.\ \cite{cll99,acl02,yl03})
modelling pulsations of superfluid neutron stars 
have been made within this approach
(see, however, \cite{ga06}).

In this paper, we will not use the Carter's hydrodynamics
because it is an essentially phenomenological theory
and it does not allow easy interpretation 
in terms of quantities calculated from microscopic theory.
Since our main goal is the calculation 
of bulk viscosity coefficients, 
we will employ the hydrodynamics of Son \cite{Son01}.
It was initially proposed in the context of heavy-ion collisions
and is derived directly from microscopic theory.
Therefore, it is straightforward to relate various parameters of 
this hydrodynamics to microphysics.
Using the notations of Ref.\ \cite{ga06} it can be rewritten
in a particularly simple form, 
which is a natural relativistic generalization of 
the standard non-relativistic superfluid hydrodynamics
pioneered by Tisza \cite{Tisza38}, 
Landau \cite{Landau41,Landau47}, 
and Khalatnikov \cite{Khalatnikov52}.

Although the Son's description is ideal 
for comparing with microphysics, 
it has some serious disadvantages. 
In contrast to the Carter's hydrodynamics, 
in which the basic fluid variables 
are the particle number density current
and the entropy density current, 
the Son's hydrodynamics is a hybrid in a sense that its 
fluid variables are the rescaled entropy density current
and the rescaled momentum of a particle 
(or a Cooper pair) from the condensate 
(in the literature it is traditionally and somewhat confusedly
referred to as `the superfluid velocity').
As a consequence, the Landau-type hydrodynamics of Son 
has a {\it lower symmetry}
than that of Carter.
However, it can be simply translated into the hydrodynamics
of Carter as was demonstrated, for example,
in the review paper by Andersson and Comer \cite{ac07} 
(see their section 16.2).

Below we will use 
the hydrodynamics of Son \cite{Son01},
employing the notations of Ref.\ \cite{ga06}, 
convenient for our problem.

Unless otherwise stated, 
the speed of light is set equal to $c=1$.

The hydrodynamic equations take the standard form
\begin{eqnarray}
\partial_{\mu} j^{\mu}=0, 
\label{hydro11} \\
\partial_{\mu} T^{\mu \nu}=0,
\label{hydro22}
\end{eqnarray}
where 
$\partial^{\mu} \equiv \partial/\partial x_{\mu}$ 
(space-time indices are denoted by Greek letters).
Neglecting dissipation, 
the particle current density $j^{\mu}$
and the energy-momentum tensor $T^{\mu \nu}$ 
can be presented in the form~\cite{ga06}
\begin{eqnarray}
j^{\mu} &=& n u^{\mu} + Y w^{\mu}, \quad
\label{current} \\
T^{\mu \nu} &=& (P+\varepsilon) \, 
u^{\mu} u^{\nu} + P \eta^{\mu \nu} 
\nonumber \\
&& + Y \left( w^{\mu} w^{\nu} + \mu \, w^{\mu} u^{\nu} 
+ \mu \, w^{\nu} u^{\mu} \right). \,\,\,\,\,
\label{Tmunu}
\end{eqnarray}
Here $n$, $P$, $\varepsilon$, and $\mu$ 
(not to be confused with the space-time index $\mu$!) 
are the number density, pressure, energy density, 
and relativistic chemical potential of particles, respectively;
$\eta^{\mu \nu}={\rm diag} (-1, +1, +1, +1)$ 
is the special-relativistic metric;
$Y$ is the relativistic analogue 
of superfluid density $\rho_{\rm s}$.
In the non-relativistic limit, we have $Y=\rho_{\rm s}/m^2$, 
where $m$ is the mass of a free particle.
Further, $u^{\mu}$ is the four-velocity 
of normal (non-superfluid) liquid component,
normalized so that
\begin{equation}
u_{\mu} u^{\mu} = -1;
\label{norma}
\end{equation}
$w^{\mu}$ is the four-velocity, 
which describes the motion of superfluid liquid component.
It can be expressed through some scalar function $\phi$,
\begin{equation}
w^{\mu}=\partial^{\mu} \phi - \mu u^{\mu}.
\label{wmu}
\end{equation}
It is easy to verify that $\phi$ is related 
to the wave function phase $\Phi$
of the Cooper-pair condensate by the equality
${\pmb \triangledown} \phi 
= (\hbar/2) \,\,\, {\pmb \triangledown} \Phi$ 
(see Ref.\ \cite{ga06}).
In the non-relativistic limit 
spatial components of the four-vectors
$u^{\mu}$ and $w^{\mu}$ are equal to
\begin{equation}
{\pmb u} = {\pmb V}_{\rm q}, \qquad
{\pmb w} = m ({\pmb V}_{\rm s} - {\pmb V}_{\rm q}),
\label{nonrel}
\end{equation}
where ${\pmb V}_{\rm s}={\pmb \triangledown} \phi /m$ 
and ${\pmb V}_{\rm q}$ are, respectively,
the superfluid and normal velocities
of the well known non-relativistic theory of superfluid liquids 
(see, e.g., Ref.\ \cite{khalatnikov89}).

The four-velocity $w^{\mu}$ must satisfy one additional constraint.
This constraint fixes a {\it comoving} frame, 
that is the frame where we measure (and define)
all the thermodynamic quantities.
Choosing the constraint in the form
\begin{equation}
u_{\mu} w^{\mu}=0,
\label{wu}
\end{equation}
from Eqs.\ (\ref{current}) and (\ref{Tmunu}) one obtains 
that in this particular case
{\it comoving} is the frame 
where four-velocity equals $u^{\mu}=(1,0,0,0)$.
It is straightforward to show that in this frame
the basic thermodynamic quantities 
$n$, ${\pmb \triangledown} \phi$, and $\varepsilon$
are defined by
\begin{eqnarray}
j^0 &=& n, \quad \quad 
\label{condition1} \\
{\pmb j} &=& Y \, {\pmb \triangledown} \phi, \quad \quad 
\label{condition2} \\
T^{00} &=& \varepsilon.
\label{condition3}
\end{eqnarray}
Using some equation of state and taking into account the second law 
of thermodynamics ($T$ is the temperature, $S$ is the entropy density)
\begin{equation}
\dd \varepsilon = T \, \dd S + \mu \, \dd n +
{ Y \over 2} \, \dd \left( w^{\mu} w_{\mu} \right),
\label{2ndlaw}
\end{equation}
as well as the definition of the pressure
\begin{equation}
P \equiv -\varepsilon + \mu n + TS,
\label{Pdefinition} 
\end{equation}
we can express all other thermodynamic quantities as functions
of $\varepsilon$,~$n$,~and~${\pmb \triangledown} \phi$.
Eqs.\ (\ref{hydro11})--(\ref{Pdefinition}) fully describe
the non-dissipative relativistic hydrodynamics 
of uncharged superfluid liquid.
As a consequence of these formulae, one can 
easily derive the continuity equation for the entropy,
\begin{equation}
\partial_{\mu}(Su^{\mu})=0.
\label{entropy}
\end{equation}

Now let us include dissipation 
in the hydrodynamics described above.
As in the non-dissipative case, 
we assume that in the comoving frame
[in which $u^{\mu}=(1,0,0,0)$], 
the basic thermodynamic quantities 
$\varepsilon$,~$n$,~and~${\pmb \triangledown} \phi$
are still defined by Eqs.\ (\ref{condition1})--(\ref{condition3}).
Being written in a relativistically invariant form, 
the conditions (\ref{condition1}) and (\ref{condition2}) imply 
that the particle current density $j^{\mu}$ 
is still given by Eq.\ (\ref{current}),
where some four-velocity $w^{\mu}$ 
satisfies the constraint (\ref{wu})
[as in the non-dissipative case].
In view of Eqs.\ (\ref{condition1}) and (\ref{condition2}),
the most general expression for the four-velocity $w^{\mu}$ is
\begin{equation}
w^{\mu} \equiv \partial^{\mu} \phi - (\mu + \varkappa) u^{\mu}.
\label{wmu1}
\end{equation}
Here a scalar $\varkappa$ is a small dissipative correction
to be determined below.
If we neglect dissipation, then $\varkappa=0$ and
Eq.\ (\ref{wmu1}) coincides naturally with (\ref{wmu}).

Without any loss of generality, 
the expression for the energy-momentum 
tensor $T^{\mu \nu}$ can be presented in the form:
\begin{eqnarray}
T^{\mu \nu} &=& (P+\varepsilon) \, u^{\mu} u^{\nu} + P \eta^{\mu \nu} 
\nonumber \\
&&+ Y \left( w^{\mu} w^{\nu} + \mu \, w^{\mu} u^{\nu} 
+ \mu \, w^{\nu} u^{\mu} \right) + \tau^{\mu \nu}.
\label{Tmunu_diss}
\end{eqnarray}
Here $\tau^{\mu \nu}$ is an unknown dissipative tensor.
In view of Eqs.\ (\ref{wu}) and (\ref{condition3}),
$\tau^{\mu \nu}$ satisfies the constraint
\begin{equation}
u_{\mu} u_{\nu} \tau^{\mu \nu} =0.
\label{tau_munu}
\end{equation}
Let us determine the dissipative corrections
$\tau^{\mu \nu}$ and $\varkappa$ assuming they are linear 
in small gradients of hydrodynamic variables.
For this aim we need to derive an entropy generation equation.
It can be easily obtained from Eqs.\ (\ref{hydro11})--(\ref{hydro22}) 
if we make use of Eqs.\ (\ref{current}), (\ref{wu}), (\ref{2ndlaw}), 
(\ref{Pdefinition}), (\ref{wmu1}), and (\ref{Tmunu_diss}),
\begin{eqnarray}
\partial_{\mu} S^{\mu} &=& 
- {\varkappa  \over T} \,\, \partial_{\mu} \left(Y w^{\mu} \right) 
-\tau^{\mu \nu} \,\, \partial_{\mu} \left( {u_{\nu} \over T} \right)
\nonumber \\
&&+ Y w^{\mu} \,\, {\varkappa  \over T^2} \,\, \partial_{\mu} T
+ u^{\nu} \,\, Y w^{\mu} \,\, {\varkappa \over T} \,\, 
\partial_{\nu} u_{\mu}.
\label{entropy1}
\end{eqnarray}
Here the entropy current density $S^{\mu}$ is given by
\begin{equation}
S^{\mu} = S u^{\mu} 
- {u_{\nu} \over T} \,\, \tau^{\mu \nu} 
- {\varkappa \over T} \,\, Y w^{\mu},
\label{S_mu}
\end{equation}
and satisfies the natural constraint
$u_{\mu} S^{\mu} =-S$.

Let us analyze the last two terms in Eq.\ (\ref{entropy1}).
In addition to a quadratic dependence on 
small gradients of hydrodynamic variables,
they also depend on the four-velocity $w^{\mu}$.
As follows from Eq.\ (\ref{nonrel}), 
the spatial part of the four-vector $w^{\mu}$ 
is proportional to the difference between 
the superfluid and normal velocities.
In a large variety of problems this difference is small 
(as a consequence, the time component $w^0$ 
is also small because of the constraint \ref{wu}). 
In particular, it cannot exceed some 
(not very large) critical value $\Delta {\pmb V}_{\rm cr}$ 
at which superfluidity breaks down (see Sec.\ \ref{4}).
For instance, if we study small perturbations of matter
which is initially at rest 
(in thermodynamic equilibrium with  $u^{\mu}=0$ and $w^{\mu}=0$), 
then the last two terms in Eq.\ (\ref{entropy1}) 
are much smaller 
than the first two terms
(a typical example of such situation 
is provided by sound waves in superfluid npe-matter, see Sec.\ \ref{5}).
Below, we will neglect the last two terms in Eq.\ (\ref{entropy1})
when obtaining the dissipative corrections
$\tau^{\mu \nu}$~and~$\varkappa$.

Moreover, 
in the dissipative corrections
we will also neglect small dissipative terms,
explicitly depending on $w^{\mu}$.
For example, we will neglect the terms of the form 
$w^{\mu} \partial_{\mu}T$ and 
$u^{\mu} w^{\nu} \partial_{\gamma} u^{\gamma}$ 
in the expressions for $\varkappa$
and $\tau^{\mu \nu}$, respectively.
An inclusion of
these small terms would result in
13 kinetic coefficients describing dissipation in superfluid liquid.
In the non-relativistic case the same approximation 
is used, for instance, in the textbook 
by Landau and Lifshitz \cite{ll87} (see their \S 140) 
and in the monograph by Khalatnikov \cite{khalatnikov89}.

Since the entropy does not decrease, 
the right-hand side of Eq.\ (\ref{entropy1})
must be positive.
This requirement puts certain restrictions 
on a general form of $\tau^{\mu \nu}$ and $\varkappa$ 
(linear in gradients).
The standard consideration (see, e.g., Ref.\ \cite{weinberg71})
shows that
\begin{eqnarray}
\tau^{\mu \nu} &=&  
- \kappa \, \left( H^{\mu \gamma} \, u^{\nu} 
+ H^{\nu \gamma} \, u^{\mu} \right) 
\left(  \partial_{\gamma} T + T u^{\delta} \,
\partial_{\delta} u_{\gamma} \right)
\nonumber \\
&-& \eta \, H^{\mu \gamma} \, H^{\nu \delta} \,\, 
\left( \partial_{\delta} u_{\gamma} + \partial_{\gamma} u_{\delta}  
- {2 \over 3} \,\, \eta_{\gamma \delta} \,\, 
\partial_{\varepsilon} u^{\varepsilon}  \right) 
\nonumber \\
&-& \xi_1 \, H^{\mu \nu} \, \partial_{\gamma} \left( Y w^{\gamma} \right) 
- \xi_2 \, H^{\mu \nu} \, \partial_{\gamma} u^{\gamma},
\label{tau_munu1} \\
\varkappa &=&  
- \xi_3 \, \partial_{\mu} \left( Y w^{\mu} \right)
- \xi_4 \, \partial_{\mu} u^{\mu}.
\label{kappa}
\end{eqnarray}
In these equations 
$H^{\mu \nu} \equiv \eta^{\mu \nu} + u^{\mu} u^{\nu}$;
$\kappa$ and $\eta$ are, respectively, the thermal conductivity 
and shear viscosity coefficients; 
$\xi_1$,$\ldots$,$\xi_4$ are the bulk viscosity coefficients.
From the Onsager symmetry principle it follows that
\begin{equation}
\xi_1=\xi_4.
\label{Onsager}
\end{equation}
For the positive definiteness of the quadratic form in the right-hand side 
of Eq.\ (\ref{entropy1}) it is necessary to have the
kinetic coefficients 
$\kappa$, $\eta$, $\xi_2$, and $\xi_3$ positive
and the coefficient $\xi_1$ satisfying the inequality
\begin{equation}
\xi_1^2 \leq \xi_2 \xi_3.
\label{noneq}
\end{equation}

In the non-relativistic limit the dissipative hydrodynamics 
proposed here coincides with the well known theory of Khalatnikov 
(see, e.g., Refs.\ \cite{khalatnikov52,khalatnikov89,ll87}).
For illustration, let us indicate how the bulk viscosity
coefficients $\xi_{\rm Kh 1}$,$\ldots$,$\xi_{\rm Kh 4}$ of Khalatnikov
are related to those introduced in this paper.
It is easy to demonstrate that 
\begin{eqnarray}
\xi_{\rm Kh1} &=& {\xi_1 \over m},  \qquad  \,\,
\xi_{\rm Kh2}=\xi_2, 
\label{relation1} \\
\xi_{\rm Kh3} &=& {\xi_3 \over m^2},  \qquad 
\xi_{\rm Kh4}  = {\xi_4 \over m}. 
\label{relation2}
\end{eqnarray}
Thus, in this section 
we have constructed the relativistic dissipative 
hydrodynamics of a superfluid liquid, 
composed of identical particles.
It should be noted, that the dissipation 
was first included into the hydrodynamics \cite{Son01} 
by Pujol and Davesne \cite{pd03}. 
However, it is difficult 
to use their dissipative hydrodynamics in applications.
The point is that the authors do not specify the {\it comoving} frame,
where they define thermodynamic quantities.
It is easy to verify that the frame which 
is defined as comoving
in our paper, cannot serve as comoving
in Ref.\ \cite{pd03}.

\section{Viscosity in superfluid mixtures}
\label{3}

Let us apply the general formulae obtained in Sec.\ \ref{2} 
to neutron star matter.
As already mentioned in Sec.\ \ref{1}, 
we consider the simplest
model of neutron star cores composed of 
neutrons (n), protons (p), and electrons (e).
All results of this and following sections can be easily generalized
to the case of matter with more complicated composition 
(e.g., npe-matter with admixture of muons or hyperon matter).

It is generally agreed that 
as a neutron star cools down
the neutrons and protons
become superfluid in its core. 
In such a system we have three velocity fields 
(instead of two, as in the previous section).
They are superfluid velocity of neutrons, 
superfluid velocity of protons,
and normal velocity $u^{\mu}$ 
of neutron and proton Bogoliubov quasiparticles
and electrons.
In this section we do not consider the dissipative effects 
related to the diffusion of particles.
Neglecting the diffusion, nucleon Bogoliubov excitations and electrons
move with the same velocity $u^{\mu}$.
In analogy with Sec.\ \ref{2}, it is convenient to introduce 
four-velocities $w^{\mu}_{(\rm n)}$ and $w^{\mu}_{(\rm p)}$
instead of superfluid velocities of neutrons and protons, respectively.
The Son's version of non-dissipative hydrodynamics 
was extended to the case of npe-mixture 
in Ref.\ \cite{ga06} 
(for earlier formulations see, e.g., 
Refs.\ \cite{Mendell91a,Mendell91b,lm94,cll99,ac01,ac06}).
The main goal of this section is to include 
the viscous dissipative terms
into the hydrodynamics \cite{ga06}. 
Below the subscripts $i$ and $k$ refer to nucleons,
$i,k={\rm n, p}$.
Unless otherwise stated, the summation is assumed over 
repeated nucleon indices $i$~and~$k$.

The full set of hydrodynamic equations 
describing superfluid mixtures
consists of
i) energy-momentum conservation law (\ref{hydro22})  
with the energy-momentum tensor $T^{\mu \nu}$ given by 
\begin{eqnarray}
&& T^{\mu \nu} = (P+\varepsilon) \, u^{\mu} u^{\nu} 
+ P \eta^{\mu \nu} 
\nonumber \\
&&+ Y_{ik} \left[ w^{\mu}_{(i)} w^{\nu}_{(k)} 
+ \mu_i \, w^{\mu}_{(k)} u^{\nu} 
+ \mu_k \, w^{\nu}_{(i)} u^{\mu} \right] + \tau^{\mu \nu}; \qquad
\label{Tmunu2}
\end{eqnarray}
ii) particle conservation laws written 
for neutrons, protons, and electrons
($l={\rm n,p,e}$),
\begin{equation}
\partial_{\mu} j^{\mu}_{ (l) } = 0, \quad 
j^{\mu}_{(i)} = n_i u^{\mu} + Y_{ik} w^{\mu}_{(k)}, \quad
j^{\mu}_{({\rm e})} = n_{\rm e} u^{\mu};
\label{particle_conservation}
\end{equation}
iii) constraints on the four-velocities $w^{\mu}_{(i)}$,
\begin{equation}
u_{\mu} w^{\mu}_{(i)} = 0,
\label{w_mu2}
\end{equation}
and iv) the second law of thermodynamics,
\begin{equation}
\dd \varepsilon = T \, \dd S + \mu_i \, 
\dd n_i +  \mu_{\rm e} \, \dd  n_{\rm e}  
+ { Y_{ik} \over 2} \, 
\dd \left[ w^{\mu}_{(i)} w_{(k) \mu} \right].
\label{2ndlaw2} 
\end{equation}
To take into account potentiality of superfluid motion, 
four-velocities $w_{(i)}^{\mu}$ 
should be expressed through some scalar functions $\phi_i$ 
and presented in the form
\begin{equation}
w^{\mu}_{(i)} = \partial^{\mu} \phi_i 
- q_i A^{\mu} - (\mu_i + \varkappa_i) u^{\mu}.
\label{wmu2}
\end{equation}
Note that one can avoid introduction of these new functions $\phi_i$
in the hydrodynamics of superfluid mixtures if 
one formulates the potentiality condition (\ref{wmu2}) 
in the equivalent way
\begin{eqnarray}
&&\partial^{\nu} \left[ w^{\mu}_{(i)} 
+q_i A^{\mu} + (\mu_i + \varkappa_i) u^{\mu} \right]
\nonumber \\
&&= \partial^{\mu} \left[ w^{\nu}_{(i)} 
 +q_i A^{\nu} +(\mu_i + \varkappa_i) u^{\nu} \right].
\label{wmu22}
\end{eqnarray}
Below we will use the latter formulation 
because it is more suitable for our purpose.
In this approach four-velocities $w^{\mu}_{(i)}$  
should be treated as independent hydrodynamic variables. 

In Eqs.\ (\ref{Tmunu2})--(\ref{wmu22}) $\mu_l$ and $n_l$
are, respectively, the relativistic chemical potential 
and the number density of particle species $l={\rm n,p,e}$;
$A^{\mu}$ is the four-potential of the electromagnetic field;
$q_i$ is the electric charge of nucleon species $i$.
Furthermore, $Y_{ik}=Y_{ki}$ 
is a $2 \times 2$ symmetric matrix 
which naturally appears in the theory  
as a generalization of the superfluid density 
to the case of superfluid mixtures.
In the non-relativistic limit this matrix is related 
to the entrainment matrix $\rho_{ik}$ 
(see Refs.\ \cite{ab75,bjk96,gh05,ga06}) by the equality
$Y_{ik}=\rho_{ik}/(m_i m_k)$, 
where $m_i$ is the mass of nucleon species $i$.
The pressure $P$ is defined in the same way as for 
non-superfluid npe-matter 
(compare with Eq.\ \ref{Pdefinition}),
\begin{equation}
P \equiv -\varepsilon + \mu_i n_i + \mu_{\rm e} n_{\rm e} + T S.
\label{Pres}
\end{equation}
The dissipative hydrodynamics formulated above
differs from the hydrodynamics \cite{ga06}, describing 
superfluid mixtures, only by the dissipative terms 
$\tau^{\mu \nu}$, $\varkappa_{\rm n}$, 
and $\varkappa_{\rm p}$.
The general form of these terms can be found
from the entropy generation equation 
which is analogous to Eq.\ (\ref{entropy1}) [see Sec.\ \ref{2}],
\begin{eqnarray}
\partial_{\mu} S^{\mu} &=& 
- {\varkappa_i  \over T} \,\, \partial_{\mu} \left[ Y_{ik} w^{\mu}_{(k)} \right]
-\tau^{\mu \nu} \,\, \partial_{\mu} \left( {u_{\nu} \over T} \right)
\nonumber \\
&&+ Y_{ik} w^{\mu}_{(k)} \,\, {\varkappa_i  \over T^2} \,\, \partial_{\mu} T
+ u^{\nu} \,\, Y_{ik} w^{\mu}_{(k)} \,\, {\varkappa_i \over T} \,\, 
\partial_{\nu} u_{\mu}, \qquad
\label{entropy2}
\end{eqnarray}
where the entropy density current $S^{\mu}$ is 
\begin{equation}
S^{\mu} = S u^{\mu} 
- {u_{\nu} \over T} \,\, \tau^{\mu \nu} 
- {\varkappa_i \over T} \,\, Y_{ik} w^{\mu}_{(k)}.
\label{S_mu2}
\end{equation}
Using the requirement that the entropy does not decrease, 
one can easily obtain 
the dissipative terms $\tau^{\mu \nu}$, 
$\varkappa_{\rm n}$, and $\varkappa_{\rm p}$ 
from Eq.\ (\ref{entropy2}), 
%
\begin{eqnarray}
\tau^{\mu \nu} &=&  
- \kappa \, \left( H^{\mu \gamma} \, u^{\nu} 
+ H^{\nu \gamma} \, u^{\mu} \right) 
\left(  \partial_{\gamma} T + T u^{\delta} \,
\partial_{\delta} u_{\gamma} \right)
\nonumber \\
&-& \eta \, H^{\mu \gamma} \, H^{\nu \delta} \,\, 
\left( \partial_{\delta} u_{\gamma} + \partial_{\gamma} u_{\delta}  
- {2 \over 3} \,\, \eta_{\gamma \delta} \,\, 
\partial_{\varepsilon} u^{\varepsilon}  \right) 
\nonumber \\
&-& \xi_{1i} \, H^{\mu \nu} \, 
\partial_{\gamma} \left[ Y_{ik} w^{\gamma}_{(k)} \right] 
- \xi_{2} \, H^{\mu \nu} \, \partial_{\gamma} u^{\gamma},
\label{tau_munu2} \\
\varkappa_{\rm n} &=&  
- \xi_{3{i}} \, \partial_{\mu} \left[ Y_{{i}k} w^{\mu}_{(k)} \right]
- \xi_{4{\rm n}} \, \partial_{\mu} u^{\mu},
\label{kappa_n} \\
\varkappa_{\rm p} &=&  
- \xi_{5{i}} \, \partial_{\mu} \left[ Y_{{i}k} w^{\mu}_{(k)} \right]
- \xi_{4 {\rm p}} \, \partial_{\mu} u^{\mu}.
\label{kappa_p}
\end{eqnarray}
Here, as in Sec.\ \ref{2}, we omit small dissipative terms, 
explicitly depending on $w^{\mu}_{(i)}$ and, in addition,
we neglect particle diffusion. 
An inclusion of these terms would result in 19 kinetic coefficients
as it has been recently demonstrated by Andersson and Comer \cite{ac06} 
for the case of npe-matter 
(they obtained their result 
using a non-relativistic version of the Carter's hydrodynamics).

In Eqs.\ (\ref{tau_munu2})--(\ref{kappa_p}) 
$\xi_{1i}$, $\xi_{2}$, $\xi_{3i}$, $\xi_{4i}$, and $\xi_{5i}$
are the bulk viscosity coefficients ($i={\rm n, p}$).
Some of them are related by the Onsager symmetry principle,
\begin{equation}
\xi_{1i}=\xi_{4i}, \qquad \xi_{3{\rm p}}=\xi_{5{\rm n}}.
\label{Onsager2}
\end{equation}
In addition, for positive definiteness of the quadratic form
in the right-hand side of Eq.\ (\ref{entropy2}) 
one needs the following inequalities
\begin{eqnarray}
&& \kappa \geq 0, \quad
\eta \geq 0, \quad 
\xi_{\rm 3n} \geq 0, \quad 
\xi_{\rm 5p} \geq 0, \quad
\xi_{\rm 2} \geq  0, 
\nonumber \\
&&\xi_{\rm 5p} \xi_{2}    \geq \xi_{\rm 1p}^2, \quad
\xi_{\rm 3n} \xi_{2}      \geq \xi_{\rm 1n}^2,  \quad
\xi_{\rm 3n} \xi_{\rm 5p} \geq \xi_{\rm 3p}^2,
\nonumber \\
&&2 \xi_{\rm 1n} \xi_{\rm 1p} \xi_{\rm 3p}
+ \xi_2 \xi_{\rm 3n} \xi_{\rm 5p}
- \xi_{\rm 1p}^2 \xi_{\rm 3n} 
- \xi_{\rm 3p}^2 \xi_{\rm 2} 
-\xi_{\rm 1n}^2 \xi_{\rm 5p} \geq 0. \qquad  
\label{noneq2}
\end{eqnarray}

Equations (\ref{entropy2})--(\ref{noneq2}) are derived
under the assumption that 
electrons and protons can move independently. 
However, this is not the case since they are charged. 
Any macroscopic motion of electrons is accompanied 
by that of protons to ensure quasineutrality condition
(see, e.g., \cite{ga06}),
\begin{equation}
n_{\rm e} = n_{\rm p}. 
\label{quasineutrality}
\end{equation}
One can obtain then from the continuity equations 
(\ref{particle_conservation})
for protons and electrons (neglecting small `diffusive' terms),
\begin{equation}
\partial_{\mu} \left[ Y_{{\rm p}k} w^{\mu}_{(k)} 
\right] = 0. 
\label{current_equality}
\end{equation}
In principle (if we are not interested
in the distribution of the electromagnetic field,
which couples together electrons and protons),
we can use this equation instead of 
the constraints (\ref{w_mu2}) and (\ref{wmu22}) for protons.

Equations (\ref{entropy2})--(\ref{noneq2}) should be modified
to take into account the conditions 
(\ref{quasineutrality}) and (\ref{current_equality}).
As follows from Eq.\ (\ref{current_equality}),
the only bulk viscosity coefficients 
which contribute  
to 
$\tau^{\mu \nu}$ and $\varkappa_{\rm n}$ 
(see Eqs.\ \ref{tau_munu2} and \ref{kappa_n})
are $\xi_{\rm 1n}$, $\xi_2$, 
$\xi_{\rm 3n}$, and $\xi_{\rm 4n}$.
Since we neglect the last two terms 
in the right-hand side 
of the entropy generation equation (\ref{entropy2}),
only these four coefficients are responsible for
dissipation of mechanical energy 
of macroscopic motion in superfluid npe-matter.

\section{Non-equilibrium beta-processes 
and the calculation of bulk viscosity}
\label{4}
In Sec.\ \ref{3} we phenomenologically considered npe-matter
and found the viscous terms 
in the relativistic hydrodynamic equations
for superfluid mixtures.
When doing this, we have ignored the fact that 
because of non-equilibrium beta-processes, 
the number of particles in the system 
is not conserved.
In this section we demonstrate
that the effect of non-equilibrium beta-processes
is equivalent to the appearance of four effective
bulk viscosity coefficients 
$\xi_{\rm 1n}$, $\xi_2$, 
$\xi_{\rm 3n}$, and $\xi_{\rm 4n}$ 
in the hydrodynamic equations.

In the npe-matter we have two types of beta-processes 
responsible for beta-equilibration. 
They are the direct Urca process and the modified Urca process.
The powerful direct Urca process is open only 
for some equations of state with large symmetry energy
(and at large enough densities, when $\pFn \leq \pFp+\pFe$, 
$\pFn$, $\pFp$, and $\pFe$ being the Fermi momenta 
of neutrons, protons, and electrons, respectively).
The direct and inverse reactions of this process 
have the form (see, e.g., Refs.\ \cite{yls99,yp04,ykgh01})
\begin{equation}
{\rm n} \rightarrow {\rm p} + {\rm e} + \bar{\nu}, \quad 
{\rm p+e} \rightarrow {\rm n}+ \nu.
\label{durca}
\end{equation}
Here $\nu$ and $\bar{\nu}$ 
stand for neutrino and antineutrino, respectively.
If the direct Urca process is forbidden by momentum conservation,
then the main mechanism of beta-equilibration 
is the modified Urca process
\cite{yls99,ykgh01,yp04}
\begin{equation}
N + {\rm n} \rightarrow N + {\rm p + e} + \bar{\nu}, \quad 
N+ {\rm p+e} \rightarrow N + {\rm n} + \nu.
\label{murca}
\end{equation}
An additional nucleon $N={\rm n, p}$ here 
is needed to take 
an excess of momentum away and open the process.

The full thermodynamic equilibrium includes
beta-equilibrium and we have \cite{ykgh01}
\begin{equation}
\delta \mu \equiv \mu_{\rm n} 
- \mu_{\rm p} - \mu_{\rm e} =0.
\label{beta_equilibrium}
\end{equation}
In this case the number of direct reactions 
in a matter element per unit time 
is equal to the number of inverse reactions.
Thus, the total number of particles of any species remains constant.
In particular, the electron generation rate $\Delta \Gamma$ 
(that is the net number of electrons generated
in beta-reactions 
in a unit volume per unit time) is zero, $\Delta \Gamma =0$.

If we perturb the system, 
the condition (\ref{beta_equilibrium}) 
will not necessarily hold ($\delta \mu \neq 0$)
and the direct and inverse
reactions will not precisely compensate 
each other ($\Delta \Gamma \neq 0$). 
In this paper we assume that the deviation 
from the equilibrium is small,
$\delta \mu \ll k_{\rm B} T$.
Then $\Delta \Gamma$ 
can be presented in the form \cite{sawyer89,hs92,hly00,hly01,hly02}
\begin{equation}
\Delta \Gamma = 
\lambda \,\, \delta \mu.
\label{dg}
\end{equation}
Here $\lambda$ is a function of various thermodynamic quantities
defined in the equilibrium state 
(e.g., of the temperature 
and the particle number densities).
For superfluid npe-matter this function was calculated by
Haensel, Levenfish, and Yakovlev \cite{hly00,hly01}.
Note, that these authors neglected the dependence of $\Delta \Gamma$
on the scalars $w^{\alpha}_{(i)} w_{(k) \alpha}$ though (in principle)
they can be non-zero in 
thermodynamic equilibrium.
However, it seems that the results of Refs.\ \cite{hly00,hly01}
are accurate as long as 
(as an example, we take the case of $i=k={\rm n}$)
\begin{equation}
{w_{\mu (\rm n)} w_{(\rm n)}^{\mu} \over m_{\rm n} } 
\sim m_{\rm n} \, \left({\pmb V}_{{\rm sn}}-{\pmb V}_{\rm q} \right)^2 
\ll k_{\rm B} T,
\label{cond}
\end{equation}
where ${\pmb V}_{{\rm sn}}$ is the superfluid velocity of neutrons.
A numerical estimate of Eq.\ (\ref{cond}) gives
\begin{equation}
| \Delta {\pmb V}_{\rm n} | 
\equiv |{\pmb V}_{{\rm sn}}-{\pmb V}_{\rm q} |
\ll 0.01 \, c \, \left({T \over 10^9 {\rm K}} \right)^{1/2}.
\label{cond2}
\end{equation}
For clarity, we introduce the velocity of light $c$ 
in this condition.
On the other hand, as follows from the Landau criterion
(see, e.g., Ref.\ \cite{lp80}), superfluidity of neutrons breaks down
if $|\Delta {\pmb V}_{\rm n}| > |\Delta {\pmb V}_{\rm cr}|$, where
\begin{equation} 
|\Delta {\pmb V}_{\rm cr}| \sim
{\Delta_{\rm n} \over \pFn} \sim {k_{\rm B} T_{c{\rm n}} \over \pFn}
\approx 0.0003 \, c \, \left({T_{c{\rm n}} \over 10^9 {\rm K}}\right) 
\left( {n_0 \over n_{\rm n}} \right) ^{1/3}.
\label{critical}
\end{equation}
Here $\Delta_{\rm n}$ is the energy gap in the neutron dispersion relation;
$T_{c{\rm n}}$ is the critical temperature 
of neutron superfluidity onset;
$n_0=0.16$ fm$^{-3}$ is the number density 
of nucleons in saturated nuclear matter.
It is worth noting that the criterion (\ref{critical}) 
is actually
an {\it upper} limit on $|\Delta {\pmb V}_{\rm cr}|$.
In reality, a superfluid state can be destroyed at much lower 
$|\Delta {\pmb V}_{\rm cr}|$ due to the formation of vortices 
in superfluid matter \cite{feynman72,putterman74}.
Comparing Eqs.\ (\ref{cond2}) and (\ref{critical})
one can see that if the neutrons are superfluid 
then at not very low temperatures
the condition (\ref{cond2}) is always justified.

To calculate the bulk viscosity coefficients 
we will use the {\it non-dissipative} hydrodynamics 
of superfluid mixtures \cite{ga06} 
(see also Sec.\ \ref{3}).
The energy-momentum tensor $T^{\mu \nu}$ 
for such a hydrodynamics is given by Eq.\ (\ref{Tmunu2}),
while the four-velocity $w^{\mu}_{(i)}$ satisfies the conditions 
(\ref{w_mu2}) and (\ref{wmu22}). 
Notice, that the dissipative components 
$\tau^{\mu \nu}$ and $\varkappa_i$ 
in Eqs.\ (\ref{Tmunu2}) and (\ref{wmu22}) should be taken zero,
$\tau^{\mu \nu}=0$ and $\varkappa_i=0$.
Further, we assume that the quasineutrality condition
(\ref{quasineutrality}) is fulfilled 
in both equilibrated and non-equilibrated matter.
Using Eq.\ (\ref{quasineutrality}), the second law
of thermodynamics for mixtures (\ref{2ndlaw2}) 
can be rewritten in the form:
\begin{equation}
\dd \varepsilon = T \, \dd S + 
\mu_{\rm n} \dd n_{\rm b} - \delta \mu \, \dd n_{\rm e}
+ { Y_{ik} \over 2} \, 
\dd \left[ w^{\mu}_{(i)} w_{(k) \mu} \right].
\label{2ndlaw3} 
\end{equation}
Finally, let us assume that
the four-velocities $w^{\mu}_{(i)}=0$ 
in equilibrium.

To take the non-equilibrium beta-processes into consideration
it is necessary to add corresponding sources in the right-hand  sides 
of the continuity equations for electrons, protons, and neutrons,
\begin{eqnarray}
&& \partial_{\mu} \left( n_{\rm e} u^{\mu} \right) 
= \Delta \Gamma,
\label{current2e} \\
&& \partial_{\mu} \left[ n_{\rm p} u^{\mu} 
+ Y_{{\rm p} k} w_{(k)}^{\mu} \right] 
= \Delta \Gamma,
\label{current2p} \\
&& \partial_{\mu} \left[ n_{\rm n} u^{\mu} 
+ Y_{{\rm n} k} w_{(k)}^{\mu} \right] 
= -\Delta \Gamma.
\label{current2n} 
\end{eqnarray}
When writing Eqs.\ (\ref{current2e})--(\ref{current2n})
we bear in mind that every neutron decay is accompanied by 
the appearance of an electron and a proton 
(see the reactions \ref{durca} and \ref{murca}).

Taking into account the quasineutrality condition (\ref{quasineutrality}), 
one gets from Eqs.\ (\ref{current2e}) and (\ref{current2p}) 
the equality (\ref{current_equality}).
Thus, Eq.\ (\ref{current_equality}) remains the same as in the absence of beta-processes.
It is more convenient to use the continuity equation for baryons
instead of Eqs.\ (\ref{current2p}) and (\ref{current2n}), 
because non-equilibrium beta-processes do not influence
the total number of baryons per unit volume, 
$n_{\rm b}=n_{\rm n}+n_{\rm p}$.
Summing together Eqs.\ (\ref{current2p}) and (\ref{current2n})
and using Eq.\ (\ref{current_equality}), one obtains
\begin{equation}
\partial_{\mu} \left[ n_{\rm b} u^{\mu} 
+ Y_{{\rm n} k} w_{(k)}^{\mu} \right] = 0.
\label{nb}
\end{equation}

Let us assume that npe-matter is slightly perturbed out 
of thermodynamic equilibrium so that 
deviations from the equilibrium are small 
and one can linearize the hydrodynamic equations.
Below we will work in the comoving 
(at one particular moment) frame
associated with some element of npe-matter.
In such a frame 
$u^{\mu}=(1,0,0,0)$.
We can assume further that
perturbations in the comoving frame depend on time $t$ as
${\rm exp}(i \omega_{\rm c} t)$, 
where $\omega_{\rm c}$ is the frequency of perturbation 
(measured in this frame).

From the normalization condition (\ref{norma}) 
and Eq.\ (\ref{w_mu2}) it follows that
\begin{eqnarray}
w^{0}_{(i)} &=& 0,
\label{w0} \\
\partial_{t} u^0 &=& 0, \quad
\partial_t w^{0}_{(i)} =0.
\label{partial0}
\end{eqnarray}
Using these equalities, one gets from 
Eqs.\ (\ref{current2e}) and (\ref{nb})
\begin{eqnarray}
&& \partial_t n_{\rm e} + {\rm div} 
\left( n_{\rm e} {\pmb u} \right) = \Delta \Gamma,
\label{ne_expand} \\
&& \partial_t n_{\rm b} + {\rm div} 
\left[ n_{\rm b} {\pmb u} + Y_{{\rm n} k} {\pmb w}_{(k)} \right] =0.
\label{nb_expand}
\end{eqnarray}
Here ${\pmb u}$ and ${\pmb w_{(k)}}$ are the spatial components
of four-vectors $u^{\mu}$ and $w^{\mu}_{(k)}$, respectively.
The number densities of electrons $n_{\rm e}$ 
and baryons $n_{\rm b}$ can be presented as
$n_{\rm e} = n_{\rm e0} + \delta n_{\rm e}$,
$n_{\rm b} = n_{\rm b0} + \delta n_{\rm b}$,
where $n_{\rm e0}$ and $n_{\rm b0}$ 
are the equilibrium number densities,
while $\delta n_{\rm e}$ and $\delta n_{\rm b}$ 
are small non-equilibrium terms
depending on time as ${\rm exp}(i \omega_{\rm c} t)$.
Here and hereafter the thermodynamic quantities 
related to the equilibrium state 
will be denoted by the subscript `0'.
Using these notations 
as well as formula (\ref{dg}) and linearizing
Eqs.\ (\ref{ne_expand})--(\ref{nb_expand}), we get
\begin{eqnarray}
\delta n_{\rm e} &=&
{1 \over i \omega_{\rm c}} \,\, 
\left[ \lambda \,\, \delta \mu 
- n_{\rm e0} \, {\rm div} ({\pmb u})
\right],
\label{ne_expand1} \\
\delta n_{\rm b} &=& -{1 \over i \omega_{\rm c}} \,\, 
\left\{ 
n_{\rm b0} \, {\rm div} ({\pmb u})
+ {\rm div} \left[ Y_{\rm nk} {\pmb w}_{(k)}\right] 
\right\}.
\label{nb_expand1}
\end{eqnarray}
Notice, that the chemical potential disbalance $\delta \mu$
in Eq.\ (\ref{ne_expand1}) depends on
$\delta n_{\rm e}$ and $\delta n_{\rm b}$.
Actually, $\delta \mu$ can generally be
presented as a function of
$n_{\rm b}$, $n_{\rm e}$, $T$, and 
the scalars $w_{\mu (i)} w^{\mu}_{(k)}$
(the proton number density is equal to the electron one, 
see the quasineutrality condition \ref{quasineutrality}).
One can neglect the temperature dependence of $\delta \mu$ 
in a strongly degenerate npe-matter
(see, e.g., Refs.\ \cite{reisenegger95,gyg05}).
Moreover, since the scalars $w_{\mu (i)} w^{\mu}_{(k)}$
are of the second order smallness,
their contribution to 
$\delta \mu$ is also negligible 
(we recall that $w^{\mu}_{(i)}=0$ in equilibrium).
Expanding
$\delta \mu(n_{\rm b}, n_{\rm e})$ in Taylor series
in the vicinity of its equilibrium value
(which is zero, $\delta \mu(n_{\rm b0}, n_{\rm e0})=0$, 
see Eq.\ \ref{beta_equilibrium}), 
one obtains in the first approximation
\begin{equation}
\delta \mu(n_{\rm b}, n_{\rm e}) = {\partial \delta 
\mu (n_{\rm b0}, n_{\rm e0}) 
\over \partial n_{\rm b0}} 
\,\, \delta n_{\rm b}
+ {\partial \delta \mu (n_{\rm b0}, n_{\rm e0}) 
\over \partial n_{\rm e0}} 
\,\, \delta n_{\rm e}.
\label{dmu_cond}
\end{equation}
Analogous formulae can be written for perturbations of pressure
$\delta P \equiv P(n_{\rm b}, n_{\rm e})-P_0$,
neutron chemical potential
$\delta \mu_{\rm n} \equiv 
\mu_{\rm n}(n_{\rm b}, n_{\rm e}) -\mu_{\rm n0}$,
and energy density
$\delta \varepsilon \equiv 
\varepsilon(n_{\rm b}, n_{\rm e})-\varepsilon_0$,
\begin{eqnarray}
\delta P &=& 
{\partial P (n_{\rm b0}, n_{\rm e0}) 
\over \partial n_{\rm b0}} 
\,\, \delta n_{\rm b}
+ {\partial P (n_{\rm b0}, n_{\rm e0}) 
\over \partial n_{\rm e0}} 
\,\, \delta n_{\rm e},
\label{P_cond} \\
\delta \mu_{\rm n} &=&
{\partial \mu_{\rm n} (n_{\rm b0}, n_{\rm e0}) 
\over \partial n_{\rm b0}} 
\,\, \delta n_{\rm b}
+ {\partial \mu_{\rm n} (n_{\rm b0}, n_{\rm e0}) 
\over \partial n_{\rm e0}} 
\,\, \delta n_{\rm e},
\label{mun_cond} \\
\delta \varepsilon &=& {\partial \varepsilon (n_{\rm b0}, n_{\rm e0}) 
\over \partial n_{\rm b0}} 
\,\, \delta n_{\rm b}.
\label{e_cond}
\end{eqnarray}
In the last equation we have neglected the term of the form
$[\partial \varepsilon (n_{\rm b0}, n_{\rm e0}) 
/ \partial n_{\rm e0}] \,\, \delta n_{\rm e}$. 
From the second 
law of thermodynamics (\ref{2ndlaw3})
we have 
$\partial \varepsilon (n_{\rm b0}, n_{\rm e0}) 
/ \partial n_{\rm e0}= -\delta \mu$.
Therefore, this term is quadratically small 
and can be omitted.

Using Eqs.\ (\ref{ne_expand1})--(\ref{dmu_cond}) 
one finds
\begin{eqnarray}
\delta n_{\rm e} &=& {1 \over F} \,\, 
\left\{  
i \, n_{\rm e0} \, \omega_{\rm c} \, \, {\rm div} ({\pmb u}) 
+ {\partial \delta \mu(n_{\rm b0}, n_{\rm e0})  \over \partial n_{\rm b0}} \, 
n_{\rm b0} \, \lambda \,\,  {\rm div} ({\pmb u}) \right.
\nonumber \\
&& \left. + {\partial \delta \mu(n_{\rm b0}, n_{\rm e0})  \over \partial n_{\rm b0}} \,
\lambda \,\,  {\rm div} \left[ Y_{{\rm n}k} {\pmb w}_{(k)} \right]
\right\},
\label{ne_expand2}
\end{eqnarray}
where $F \equiv \omega_{\rm c}^2 + i \, \omega_{\rm c} 
\, \lambda \, \partial \delta \mu(n_{\rm b0}, n_{\rm e0})
/\partial n_{\rm e0}$. 
For non-equilibrium Urca-processes and typical
(for neutron stars)
pulsation frequencies $\omega_{\rm c} \sim 10^3-10^4$ s$^{-1}$, 
we have (see, e.g., Refs.\ \cite{hly00,hly01})
$\omega_{\rm c} \gg \lambda \,
|\partial \delta \mu(n_{\rm b0}, n_{\rm e0})
/\partial n_{\rm e0}|$. 
In this case Eq.\ (\ref{ne_expand2}) 
can be 
simplified 
by keeping only terms linear in $\lambda$.
The result can be written as
\begin{equation}
\delta n_{\rm e} =
\delta n_{\rm e1} + \delta n_{\rm e2},
\label{ne_expand3}
\end{equation}
where the first term equals
\begin{equation}
\delta n_{\rm e1} = {i n_{\rm e0} 
\over \omega_{\rm c}} \, {\rm div}({\pmb u})
\label{dne0}
\end{equation}
and describes compression and decompression 
of the pulsating matter. 
This term remains the same even 
in the absence of non-equilibrium beta-processes.
The second term $\delta n_{\rm e2}$ 
is due to non-equilibrium beta-processes
\begin{eqnarray}
\delta n_{\rm e2} &=& {\lambda \over \omega_{\rm c}^2} \, 
\left\{
n_{\rm b0} \, {\partial \delta \mu(n_{\rm b0}, x_{\rm e0}) 
\over \partial n_{\rm b0}} \,\, {\rm div}({\pmb u}) \right.
\nonumber \\
&& \left. + {\partial \delta \mu(n_{\rm b0}, n_{\rm e0}) 
\over \partial n_{\rm b0}} \,\,
{\rm div}\left[ Y_{{\rm n}k} {\pmb w}_{(k)} \right]
\right\}.
\label{dne1}
\end{eqnarray}
Notice, that in this formula the partial derivative
$\partial \delta \mu(n_{\rm b0}, x_{\rm e0}) 
/\partial n_{\rm b0}$
is taken at constant value of
$x_{\rm e0} \equiv n_{\rm e0}/n_{\rm b0}$.
When obtaining Eq.\ (\ref{dne1}) we used the identity 
\begin{equation}
n_{\rm b0} \, {\partial \Psi(n_{\rm b0}, x_{\rm e0}) 
\over \partial n_{\rm b0}   } =
n_{\rm b0} \, {\partial \Psi(n_{\rm b0}, n_{\rm e0}) 
\over \partial n_{\rm b0}}
+n_{\rm e0} \, {\partial \Psi(n_{\rm b0}, n_{\rm e0}) 
\over \partial n_{\rm e0}},
\label{Psi}
\end{equation}
where $\Psi$ is an arbitrary function 
of $n_{\rm b0}$ and $n_{\rm e0}$.

Our further strategy is as follows.
We take the energy-momentum tensor $T^{\mu \nu}$ 
for superfluid mixtures 
(with $\tau^{\mu \nu}$=0) from Eq.\ (\ref{Tmunu2}) 
and expand all the thermodynamic quantities 
(e.g., the pressure $P$ and the energy density $\varepsilon$),
which determine this tensor, around their equilibrium values.
Restricting ourselves to linear perturbation terms,
we obtain for the tensor $T^{\mu \nu}$
(in the comoving frame),
\begin{eqnarray}
&& T^{00} = \varepsilon_0 + \delta \varepsilon,
\nonumber \\
&& T^{0j}= T^{j0} = 
\mu_{i0} Y_{ik} \, w^{j}_{(k)},
\nonumber \\
&& T^{jm} = \left( P_0
+ \delta P \right) \,\, \delta_{jm}.
\label{Tmunu3} 
\end{eqnarray}
Here, the spatial indices $j$ and $m$ are equal to $1,2,3$;
the relativistic entrainment matrix $Y_{ik}$ 
is taken in equilibrium;
$\delta P$ and $\delta \varepsilon$ 
are given by Eqs.\ (\ref{P_cond}) and (\ref{e_cond}), respectively.

Let us assume for a while that there are no 
non-equilibrium beta-processes in the matter.
In this case $\delta n_{\rm e2}=0$ (see Eq.\ \ref{dne1}) 
and the matter is reversibly pulsating
around the equilibrium.
Then the mechanical energy is not dissipating,
and the entropy is conserved.
Thus it is obvious that dissipative are only 
those terms in the tensor $T^{\mu \nu}$ 
which are directly related 
to $\delta n_{\rm e2}$. 
Writing out these terms in the form of a separate tensor
$\tau^{\mu \nu}_{\rm bulk}$, we have
\begin{eqnarray}
&& \tau^{00}_{\rm bulk} =0,
\nonumber \\
&& \tau^{0j}_{\rm bulk}=\tau^{j0}_{\rm bulk}=0,
\nonumber \\
&& \tau^{jm}_{\rm bulk}={\partial P(n_{\rm b0}, n_{\rm e0}) 
\over \partial n_{\rm e0} } \, \delta n_{\rm e2} \,\, \delta_{jm}.
\label{tau_bulk}
\end{eqnarray}
This tensor can be easily rewritten 
in an arbitrary frame if we take into account Eq.\ (\ref{dne1}),
\begin{eqnarray} 
&& \tau^{\mu \nu}_{\rm bulk} = {\lambda \over \omega_{\rm c}^2} \,
{\partial P(n_{\rm b0}, n_{\rm e0}) \over \partial n_{\rm e0} }
 \,\, H^{\mu \nu} 
\nonumber \\
&& \times \, \left\{
{\partial \delta \mu(n_{\rm b0}, n_{\rm e0}) \over \partial n_{\rm b0} }
\,\,  \partial_{\gamma} 
\left[ Y_{{\rm n}k} w^{\gamma}_{(k)} \right] \right.
\nonumber \\
&& \left. + n_{\rm b0} \, \,
{\partial \delta \mu(n_{\rm b0}, x_{\rm e0}) 
\over \partial n_{\rm b0} } \,\, \partial_{\gamma}  u^{\gamma}
\right\}.
\label{tau_bulk2}
\end{eqnarray}
Comparing the tensor $\tau^{\mu \nu}_{\rm bulk}$ with 
the phenomenological dissipative tensor $\tau^{\mu \nu}$ 
(see Eq.\ \ref{tau_munu2}), 
we find the expressions for 
the effective bulk viscosity coefficients $\xi_{\rm 1n}$ and $\xi_2$,
generated by non-equilibrium beta-processes
\begin{eqnarray}
\xi_{\rm 1n} &=& - {\lambda \over \omega_{\rm c}^2} \,\,
{\partial P(n_{\rm b0}, n_{\rm e0}) \over \partial n_{\rm e0} } \,\,
{\partial \delta \mu(n_{\rm b0}, n_{\rm e0}) \over \partial n_{\rm b0} },
\label{bulkvisc_1n} \\
\xi_{2} &=& - {\lambda \over \omega_{\rm c}^2} \,\, n_{\rm b0} \,\, 
{\partial P(n_{\rm b0}, n_{\rm e0}) \over \partial n_{\rm e0} } \,\,
{\partial \delta \mu(n_{\rm b0}, x_{\rm e0}) 
\over \partial n_{\rm b0} }.
\label{bulkvisc2}
\end{eqnarray}

Let us do the same with
the potentiality condition (\ref{wmu22}) 
on the four-velocity of neutrons $w^{\mu}_{(\rm n)}$. 
As a result, we obtain in the comoving frame
the dissipative component $\varkappa_{\rm n}$, 
appearing because of non-equilibrium beta-processes,
\begin{equation}
\varkappa_{\rm n} = {\partial \mu_{\rm n}(n_{\rm b0}, n_{\rm e0}) 
\over \partial n_{\rm e0} } \,\, \delta n_{\rm e2}.
\label{kappa_n2}
\end{equation}
In a fully covariant form, this component is given by Eq.\ (\ref{kappa_n})
where the effective bulk viscosity coefficients 
$\xi_{\rm 3n}$ and $\xi_{\rm 4n}$ are
\begin{eqnarray}
\xi_{\rm 3n} &=& - {\lambda \over \omega_{\rm c}^2} \,\,
{\partial \mu_{\rm n}(n_{\rm b0}, n_{\rm e0}) 
\over \partial n_{\rm e0} } \,\,
{\partial \delta \mu(n_{\rm b0}, n_{\rm e0}) \over \partial n_{\rm b0} },
\label{bulkvisc_3n} \\
\xi_{\rm 4n} &=& - {\lambda \over \omega_{\rm c}^2} \,\, n_{\rm b0} \,\, 
{\partial \mu_{\rm n}(n_{\rm b0}, n_{\rm e0}) 
\over \partial n_{\rm e0} } \,\,
{\partial \delta \mu(n_{\rm b0}, x_{\rm e0}) 
\over \partial n_{\rm b0} },
\label{bulkvisc_4n}
\end{eqnarray}
and, in addition, the condition (\ref{current_equality}) 
is taken into account.

Thus, we have calculated 
the four effective bulk viscosity coefficients
$\xi_{\rm 1n}$, $\xi_2$, 
$\xi_{\rm 3n}$, and $\xi_{\rm 4n}$.
As will be shown in Sec.\ \ref{5}, 
each of them 
makes a comparable contribution 
to characteristic damping times of mechanical energy.
Notice, that only the coefficient $\xi_2$ is usually analyzed
in the literature devoted to non-equilibrium beta-processes 
in superfluid matter.
The expression (\ref{bulkvisc2}) for $\xi_2$ coincides with 
earlier results (see, e.g., Refs.\ \cite{hly00,hly01}).

Not all of the coefficients (\ref{bulkvisc_1n})--(\ref{bulkvisc2}) 
and (\ref{bulkvisc_3n})--(\ref{bulkvisc_4n}) are independent.
The coefficients 
$\xi_{\rm 1n}$ and $\xi_{\rm 4n}$ 
are equal because of the Onsager principle (\ref{Onsager2}).
This can be shown if one 
applies
the following relation
for npe-matter
(see, e.g., Ref.\ \cite{hly00}),
\begin{equation}
{\partial P(n_{\rm b0}, x_{\rm e0}) 
\over \partial x_{\rm e0} } = -n_{\rm b0}^2 \,\, 
{\partial \delta \mu(n_{\rm b0}, x_{\rm e0}) 
\over \partial n_{\rm b0} }.
\label{thermodynamic}
\end{equation}
Furthermore, it is easy to verify, 
that instead of one of the inequalities (\ref{noneq2}) 
relating the coefficients 
$\xi_{\rm 1n}$, $\xi_{\rm 2}$, and $\xi_{\rm 3n}$, 
we have the strict equality
\begin{equation}
\xi_{\rm 1n}^2 =\xi_2 \xi_{\rm 3n}.
\label{xi_eq}
\end{equation}
It is fulfilled only for those non-equilibrium processes,
for which the expansion (\ref{ne_expand3}) is valid.
Therefore, we have only {\it two} 
independent bulk viscosity coefficients.

To prove Eq.\ (\ref{xi_eq})
it is instructive to consider the entropy generation equation.
Neglecting all the dissipative processes 
except for the non-equilibrium beta-processes 
(e.g., neglecting thermal conductivity, 
diffusion, shear viscosity), 
one can obtain from the hydrodynamics discussed in this section
\begin{equation}
T \, \partial_{\mu} S^{\mu} = \delta \mu \, \Delta \Gamma
=  \lambda \, \delta \mu^2.
\label{entropy3}
\end{equation}
Here we made use of Eq.\ (\ref{dg}). 
Since we are interested only in terms linear in $\lambda$,
we can substitute $\delta n_{\rm e1}$ for $\delta n_{\rm e}$ 
into Eq.\ (\ref{dmu_cond}) which determines $\delta \mu$. 

On the other hand, 
the entropy generation equation
in terms of the effective bulk viscosities
takes the form 
(see Eqs.\ \ref{entropy2}, (\ref{tau_munu2})--(\ref{kappa_p}), 
and \ref{current_equality}),
\begin{eqnarray}
&&T \, \partial_{\mu} S^{\mu} = \xi_{\rm 3n}  \, 
\left\{ 
\partial_{\mu} \left[ 
Y_{{\rm n}k} w^{\mu}_{(k)} 
\right] 
\right\}^2 
\nonumber \\
&&+ (\xi_{\rm 1n}+\xi_{\rm 4n}) \,\, \partial_{\mu} \left[ 
Y_{{\rm n}k} w^{\mu}_{(k)} \right] \, \partial_{\mu} u^{\mu}
+ \xi_2 \, \left( \partial_{\mu} u^{\mu} \right)^2. \qquad \qquad
\label{entropy4}
\end{eqnarray}

Let us compare the right-hand sides of 
Eqs.\ (\ref{entropy3}) and (\ref{entropy4}).
It follows from Eqs.\ (\ref{dmu_cond}) and (\ref{entropy3}) 
that for any given $u^{\mu}$ 
it is always possible to choose four-velocities $w^{\mu}_{(k)}$ 
in such a way, that $\delta \mu =0$ 
and the entropy generation rate vanishes
(at some point and at some particular moment).
In terms of the bulk viscosity formalism this means 
that one can vanish the quadratic form in the 
right-hand side of Eq.\ (\ref{entropy4})
by an appropriate choice of these velocities. 
This is possible {\it only} 
if the equality (\ref{xi_eq}) is satisfied.

\section{Damping of sound waves in superfluid npe-matter}
\label{5}
Let us illustrate the results of previous sections 
by calculating characteristic damping times of sound waves
propagating in a homogeneous superfluid npe-matter.
For simplicity, we consider only the damping 
due to the effective bulk viscosity.
Neglecting dissipation, 
the sound modes of superfluid npe-matter 
have been thoroughly investigated starting from the 
pioneering paper by Epstein \cite{Epstein88}
in which he argued that there would be 
two types of sound modes in neutron stars
(see, e.g., Refs.\ \cite{lm94,ac01,ga06}).
In particular, Gusakov and Andersson \cite{ga06} 
were the first who considered 
in full relativity sound modes in npe-matter 
at finite temperatures. 
Here we closely follow their analysis.
The pulsation equations (81) and (82) of Ref.\ \cite{ga06}
can be used to describe sound waves
taking into account dissipation.
Thus, there is no need to derive these equations 
from the hydrodynamics of superfluid mixtures 
(Secs.\ \ref{3}~and~\ref{4}) once again.
Instead, we will rewrite them using the notations adopted in our paper.
The result is
\begin{eqnarray}
&& \partial_t 
\left[ 
(P_0 + \varepsilon_0) \, {\pmb u}  
+ \mu_{\rm n0} Y_{{\rm n}k} \, {\pmb w}_{(k)}
\right] 
= -{\pmb \triangledown} \delta P,
\label{sound1} \\
&&\partial_t 
\left[ 
\mu_{\rm n0} \, {\pmb u} + {\pmb w}_{(\rm n)}
\right] 
= - {\pmb \triangledown} \delta \mu_{\rm n}.
\label{sound2} 
\end{eqnarray}
The first equation is a consequence of the relativistic Euler equation,
which can be derived from Eq.\ (\ref{hydro22}) in a standard way
(see, e.g., Ref.\ \cite{ll87}).
The second equation follows from 
the condition (\ref{wmu22}) written for neutrons. 
To fully define the system, Eqs.\ (\ref{sound1}) and (\ref{sound2}) 
should be supplemented by the condition (\ref{current_equality}).
Using Eqs.\ (\ref{w0}) and (\ref{partial0}), 
this condition can be presented in the form
\begin{equation}
{\rm div} \left[ Y_{{\rm p} k} {\pmb w}_{(k)}\right]=0.
\label{current_equality1}
\end{equation}
Now assuming that all the perturbations are plane waves
proportional to ${\rm exp}(i \omega t - i{\pmb k}{\pmb r})$,
one obtains the following compatibility condition for 
Eqs.\ (\ref{sound1})--(\ref{current_equality1})
[$s = \omega / k$ is the velocity of sound in units of $c$]
\begin{equation}
y \, s^4 + C_1 s^2 + C_2 + \delta A =0, 
\label{bi2}     
\end{equation}
where
\begin{eqnarray}
y &=& {Y_{\rm pp} \, n_{\rm b0} \over \mu_{\rm n0} \, 
\left(Y_{\rm nn} Y_{\rm pp} - Y_{\rm np} Y_{\rm pn} \right)} -1,
\label{y} \\
C_1 &=& \left[ {P_0 \over \mu_{\rm n0} n_{\rm b0}} \,\left(
\beta_1 - \gamma_1 - \gamma_1 y \right) 
+ \gamma_2-\beta_2   \right],
\label{C1} \\
C_2 &=& {P_0 \over \mu_{\rm n0} n_{\rm b0}} \, 
\left( \beta_2 \gamma_1 - \beta_1 \gamma_2 \right), 
\label{C2} \\
\gamma_1 &=& {n_{\rm b0} \over P_0} \, 
{\partial P(n_{\rm b0}, x_{\rm e0}) \over \partial n_{\rm b0}},\quad 
\gamma_2 = {n_{\rm b0} \over \mu_{\rm n0}} \, 
{\partial \mu_{\rm n}(n_{\rm b0}, x_{\rm e0}) 
\over \partial n_{\rm b0}}, \qquad
\label{gamma} \\
\beta_1 &=& {n_{\rm b0} \over P_0} \, 
{\partial P(n_{\rm b0}, n_{\rm e0}) \over \partial n_{\rm b0}},\quad 
\beta_2 = {n_{\rm b0} \over \mu_{\rm n0}} \, 
{\partial \mu_{\rm n}(n_{\rm b0}, n_{\rm e0}) 
\over \partial n_{\rm b0}}. \qquad
\label{beta1} 
\end{eqnarray}
A small complex term $\delta A$  
appears in the compatibility condition (\ref{bi2}) 
because of the bulk viscosity.
It is given by
\begin{eqnarray}
 \delta A &=& -{i \omega \over \mu_{\rm n0}^2 \, n_{\rm b0}} \,
\left(  A_1 + s^2 A_2 \right), 
\label{dA} \\
A_1 &=&
\mu_{\rm n0} n_{\rm b0} \, \gamma_2  \, \xi_{\rm 1n}
- \mu_{\rm n0} \, \beta_2 \, \xi_{2}         
\nonumber \\
&&- P_0 n_{\rm b0} \, \gamma_1 \, \xi_{\rm 3n}
+P_0 \, \beta_1 \, \xi_{\rm 4n},
\label{A1} \\
A_2 &=& 
\mu_{\rm n0} \, 
\left( 
y \, \xi_2  + \xi_2 + n_{\rm b0}^2 \, \xi_{\rm 3n}
- n_{\rm b0} \, \xi_{\rm 1n} - n_{\rm b0} \, \xi_{\rm 4n} \right). \,\,\,\,\,\,\,\,\,
\label{A2}
\end{eqnarray}
We remind that the bulk viscosity coefficients 
(and the quantities $A_1$ and $A_2$) 
depend on the frequency $\omega$,
$A_{1,2} \sim \omega^{-2}$.
The biquadratic equation (\ref{bi2}) has two non-trivial 
solutions for two possible sound velocities.
Neglecting dissipation, 
these modes have been analyzed in details
in Ref.\ \cite{ga06}.
In particular, the sound velocities
$s^{(0)}_{1}$ and $s^{(0)}_{2}$ 
have been calculated there
for the first and second modes.
The dissipation leads to the appearance of small 
complex corrections $\delta s_{1,2}$ 
to the velocities $s_{1,2}^{(0)}$
and consequently to decrements of sound waves.
Since $\delta A$ is small 
in comparison with other terms in Eq.\ (\ref{bi2}), 
one can use the perturbation theory in deriving 
the characteristic damping times $\tau_{1,2}$.
The parameters $\tau_{1}$ and $\tau_{2}$ 
are e-folding times of the pulsation amplitude
for the first and second sound modes, respectively,
%
%
%
\begin{equation}
\tau_{1,2} \approx {i \over k \,\, \delta s_{1,2} }
= - {2 \, i \, s_{1,2}^{(0)} \over k \, \delta A} \, 
\left( 2\, y \, s^{(0)2}_{1,2} + C_1 \right).
\label{tau}
\end{equation}
As follows from Eqs.\ (\ref{dA})--(\ref{A2}), 
they are {\it independent}~of~$\omega$.

At $T \rightarrow T_{c {\rm n}}$ we have
$Y_{\rm nn}, Y_{\rm np},  Y_{\rm pn} \rightarrow 0$
and $y \approx n_{\rm b0}/(\mu_{\rm n0} Y_{\rm nn}) \rightarrow \infty$ 
(see Ref.\ \cite{ga06} for a more detailed discussion).
In this limit the characteristic damping times are
\begin{eqnarray}
 \tau_1 &\approx& {2 \, P_0 \, \gamma_1 \over \omega^2 \xi_2},
\label{tau1} \\
\tau_2 &\approx& -{2 \, \mu_{\rm n0} P_0 \, \gamma_1 \, D
\over  \omega^2 \, 
\left( \gamma_1 \, A_1 + \mu_{\rm n0} \, D \, \xi_2 \right)}.
\label{tau2}
\end{eqnarray}
Here we introduce the parameter
$D \equiv \beta_2 \gamma_1 - \beta_1 \gamma_2$.
At $T > T_{c {\rm n}}$ the neutrons are non-superfluid.
In this case the second mode does not exist
[formally, $s_2^{(0)}=0$], 
while the first mode is the usual sound wave.
The characteristic damping time for an ordinary sound wave
is given by Eq.\ (\ref{tau1}).
As expected, its damping 
is governed by the only one 
bulk viscosity coefficient $\xi_2$.

For illustration, in Fig. 1 we present
the characteristic damping times $\tau_{1,2}$ of sound waves (in years)
as a function of temperature $T$ for two sound modes.
The figure is plotted for npe-matter with 
the baryon number density
$n_{\rm b0}=3 n_0$.
The critical temperature of neutrons is taken to be
$T_{c {\rm n}}=10^9$ K.
The protons are assumed to be non-superfluid.
When calculating thermodynamic quantities and their derivatives 
we employed the equation of state from Ref.\ \cite{hh99}. 
It opens the direct Urca process 
at baryon number density of $5.84 n_0$.
Therefore, the process is forbidden for $n_{\rm b0}=3 n_0$.
In this case, the main mechanism of energy dissipation 
is the non-equilibrium modified Urca process.
To calculate the function $\lambda$, 
which enters Eqs.\ (\ref{bulkvisc_1n})--(\ref{bulkvisc2}) 
and (\ref{bulkvisc_3n})--(\ref{bulkvisc_4n})
for the bulk viscosity coefficients,
we have used the results of Ref.\ \cite{hly01}.
For calculating the relativistic entrainment matrix $Y_{ik}$
we have employed the BJ~v6 nucleon-nucleon potential \cite{jkms82,gh05,ga06}.
Actually, the microphysics input we use here to plot the figure
is taken from Ref.\ \cite{ga06} (see this reference for more details).
%
\begin{figure}[t]
\setlength{\unitlength}{1mm}
\leavevmode
\hskip  0mm
\includegraphics[width=85mm,bb=18  145  562  690,clip]{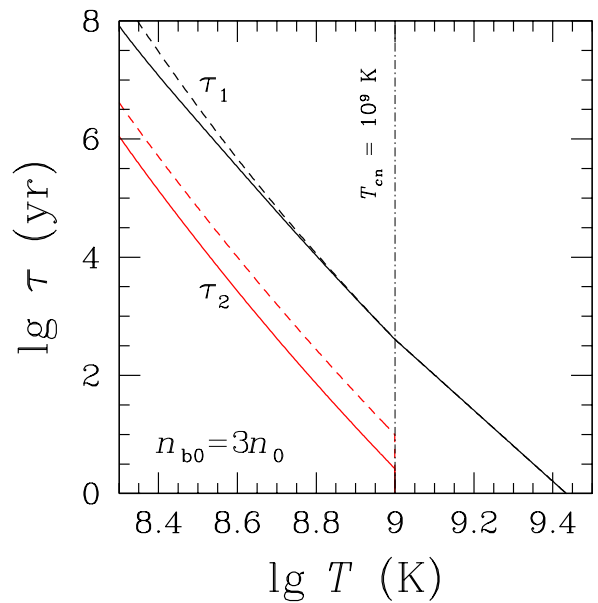}
\caption
{Characteristic damping times $\tau_{1,2}$ 
of sound waves versus temperature $T$
for the first mode (two upper curves)
and for the second mode (two lower curves).
The solid curves demonstrate the damping of sound 
taking into account 
all four bulk viscosity coefficients.
The dashed curves are calculated 
assuming that
only $\xi_2$ is non-zero.
The neutron critical temperature is indicated 
by the vertical dot-dashed line.
The baryon number density is
$n_{\rm b}=3 n_0=0.48$ fm$^{-3}$.}
\label{1fig}   
\end{figure}
%

The two upper curves correspond to the first sound mode,
while the two lower curves -- to the second mode.
The solid curves are plotted taking into account 
all four bulk viscosity coefficients.
The dashed curves are obtained under the assumption 
that all coefficients but $\xi_2$ are equal to zero,
$\xi_{\rm 1n}=\xi_{\rm 3n}=\xi_{\rm 4n} \equiv 0$.

As seen from the figure, 
the characteristic damping times of sound waves
increase as the temperature decreases.
This is natural, because when the neutrons are superfluid,
the Urca processes (and hence the bulk viscosity)
are exponentially suppressed at $T \ll T_{c{\rm n}}$.
Let us emphasize that at temperatures $T \la 5 \times 10^8$ K,
the shear viscosity of electrons can exceed 
the bulk viscosity generated 
by the non-equilibrium modified Urca process.
As a result, the damping of sound waves will be
mainly due to the shear viscosity. 

The first mode turns into the ordinary sound at $T>T_{c{\rm n}}$.
As follows from the figure, 
in the vicinity of neutron critical temperature
the dissipation is primarily determined
by the bulk viscosity coefficient $\xi_2$
(in accordance with Eq.\ \ref{tau1}).
Consequently, near the transition point 
the solid and the dashed curves 
for the first mode coincide.
On the contrary, the difference between 
the solid and the dashed curves for the second mode 
remains significant at any $T<T_{c{\rm n}}$.
The characteristic damping times for these two curves 
differ approximately by a factor of 3.

It is worth noting that we would come to the similar conclusions
if we considered sound waves in denser matter, 
where the direct Urca process is open.
In that case the characteristic damping times 
would be 
6--7 orders of magnitude smaller, 
but the relative difference between the solid 
and the dashed curves 
will be approximately the same.

Therefore, the main result of the present section 
is that the bulk viscosity coefficients 
$\xi_{\rm 1n}$, $\xi_{\rm 3n}$, and $\xi_{\rm 4n}$ 
essentially influence the dissipative properties 
of superfluid npe-matter 
and cannot be ignored.
All four bulk viscosity coefficients should be 
considered on the same footing.

\section{Summary}
\label{6}

We performed a self-consistent analysis
of the influence of non-equilibrium beta-processes
on dissipation of mechanical energy 
in superfluid matter of neutron stars. 
We start with the Son's version 
of non-dissipative one-fluid relativistic hydrodynamics 
to describe superfluid mixtures (see Refs.\ \cite{Son01,ga06}).
We determined the general form of dissipative terms 
entering the equations of this hydrodynamics.
For simplicity, the effects of particle diffusion were ignored.
The equations of dissipative hydrodynamics
were applied to the matter composed of 
neutrons, protons, and electrons (npe-matter). 
In this case the hydrodynamic equations 
contain {\it four} bulk viscosity coefficients 
rather than one, 
as in non-superfluid matter.

It was demonstrated, that non-equilibrium beta-processes
generate all four bulk viscosity coefficients, 
and only two of them are independent.
The other two coefficients can be expressed through the first two
by Eqs.\ (\ref{Onsager2}) and (\ref{xi_eq}).
It is worth to emphasize that only the bulk viscosity coefficient $\xi_2$
has been considered in the astrophysical literature so far.
The expression (\ref{bulkvisc2}) for $\xi_2$
coincides 
with similar expressions of previous works
(see, e.g., Refs.\ \cite{sawyer89,hs92,hly00,hly01}).

To illustrate the results obtained in the present paper
we considered a problem of damping of sound waves 
via the bulk viscosity due to non-equilibrium beta-processes
in superfluid homogeneous npe-matter.
It was shown that all four bulk viscosity coefficients 
make comparable contributions 
to the characteristic damping times of sound waves.

Our results
can be important for the analysis 
of various gravitational-driven instabilities
in neutron stars, in particular, the r-mode instability 
(see, e.g., Ref.\ \cite{andersson03}). 
The additional bulk viscosity coefficients lead to a more effective
damping of these instabilities.
Moreover, the results can be applied to the problems 
of rotochemical and gravitochemical heating of millisecond pulsars 
with superfluid cores. 
In the absence of superfluidity these problems 
were carefully analyzed
in Refs.\ \cite{reisenegger95,fr05,rjfk06,jrf06}.
The first attempt to discuss qualitatively
the effects of superfluidity has been made 
in Ref.\ \cite{reisenegger97}.

In conclusion let us note
that the method of the bulk viscosity calculation, 
used here in the simple case of npe-matter, 
can be extended to matter with more complicated composition 
(npe-matter with admixture of muons, hyperon or quark matter). 
Such a generalization is beyond the scope of the present study
and will be considered elsewhere.

\section*{Acknowledgments}
The author is grateful to 
D.P. Barsukov and E.M. Kantor for discussions, 
to A.I. Chugunov for technical assistance, 
to D.G. Yakovlev for reading the manuscript and critical comments, 
and to A. Reisenegger and anonymous referee 
for very useful remarks. 
This research was supported
by RFBR (grants 05-02-16245 and 05-02-22003)
and by the Federal Agency for Science and Innovations
(grant NSh 9879.2006.2).


\end{document}